\documentclass[12pt,prd,nofootinbib,superscriptaddress]{revtex4}

\usepackage{amsmath}
\usepackage{amsfonts}

\begin{document}
\title{Internal Relativity}
\author{Olaf DREYER}
\affiliation{Dipartimento di Fisica, Universit\`a di Roma ``La Sapienza"\\
and Sez.~Roma1 INFN, P.le A. Moro 2, 00185 Roma, Italy}

\begin{abstract}General relativity differs from other forces in nature in that it can be made to disappear locally.  This is the essence of the equivalence principle.  In general relativity the equivalence principle is implemented using differential geometry.  The connection that comes from a metric is used to glue together the different gravity-free Minkowski spaces.  In this article we argue that there is another way to implement the equivalence principle.  In this new way it is not different Minkowski spaces that are connected but different vacua of an underlying solid-state like model.  One advantage of this approach to gravity is that one can start with a quantum mechanical model so that the question of how to arrive at a quantum theory of gravity does not arise.  We show how the gravitational constant can be calculated in this setup. 
\end{abstract}

\maketitle

\tableofcontents

\newpage

\newpage

\section{Introduction}

What is quantum gravity?  For many quantum gravity is the quantization of the classical theory of gravity.  The process of quantization has been very successful for simple system of a small number of degrees of freedom but it also has worked for field theories with infinitely many degrees of freedom.  All the quantum field theories that so successfully describe microscopic physics start from classical Lagrangians that are then quantized.  A theory that has so far resisted attempts of this kind is general relativity although progress has been made on this front (see \cite{pullin, thieman, rovelli, loll}).  It might be that we just have not figured out the correct way of performing the quantization.  It might be a technical problem that has so far prevented us from obtaining a quantum theory of gravity via a process of quantization.  On the other hand it might be that the failure to quantize the theory is a hint that quantization is not the path to a quantum theory of gravity. 

In condensed matter physics a different point of view of the relation between a quantum theory and its classical limit has emerged.  It may happen that the degrees of freedom that describe the low energy behavior of a quantum theory are very different from the more fundamental building blocks of the theory.  Spin waves for example are very different from the underlying spins that give rise to them.  In this case the quantum theory can not be obtained by quantizing the theory of the low energy degrees of freedom.  What if the quantum theory of gravity is of this kind?  This would mean that we should not start by trying to construct a quantum theory of spacetimes.  The underlying theory would not be a theory of quantized metrics.  Instead a spacetime would only appear in the appropriate low-energy limit.  In this paper we will argue for just such a scenario. 

Even if we adopt this point of view there is one more choice that we have to make.  We have to answer the question of how gravity arises.  There are two possibilities.  The first possibility is that gravity arises as a low energy excitation of the theory.  In this possibility gravity is perturbative.  We would look for a particle that mediates gravity,  i.e. the graviton,  a zero mass spin two excitation,  together with its perturbation theory.  The other possibility is that gravity is not perturbative.  We will argue in this paper that gravity is of this second kind.  

In the next section we review what we consider to be the most important feature of gravity,  namely the equivalence principle.  Gravity is different from other forces because locally it can not be detected.  We review how this feature is implemented in general relativity.  We then argue that there is a new way of implementing the equivalence principle that does not make use of a metric as general relativity does.  We then show how this new look at gravity can be used to calculate the gravitational constant $G$.  We finish with a discussion and outlook.  

\section{General relativity and the equivalence principle}

Gravity is different from other forces in that locally it can be made to disappear.  No local measurement can tell a freely falling observer whether there is a gravitational field or not.  This is the equivalence principle.  Locally all physics is as it is in Minkowski space.  Gravity is in the way these local Minkowski spaces are connected.  

In general relativity this is implemented as follows.  A connection $\Gamma_{ab}{}^{c}$ contains the information of how to get from one local Minkowski space to another.  If $v^a$ is a vector that we want to transport to another point on a path with tangent vector $w^b$ then the change of $v^a$ is given by 
\begin{equation}
\Gamma_{ab}{}^c v^a w^b.
\end{equation}
Given a metric $g_{ab}$ we obtain a connection $\Gamma_{ab}{}^c$ through
\begin{equation}
\Gamma_{ab}{}^c = \frac{1}{2} g^{cd}(g_{db,a} + g_{ad,b} - g_{ab,d}).
\end{equation}
Through this connection the metric $g_{ab}$ becomes the glue that connects the local Minkowski spaces.  It is the new dynamical variable of the theory.

The behavior of the metric is now tied to the behavior of matter via Einstein's equations:
\begin{equation}
R_{ab} - \frac{1}{2}g_{ab}R = 8\pi  G T_{ab}
\end{equation}
In this language of metrics and connections the equivalence principle is expressed through the fact that one can always choose a coordinate system containing a point $p$ such that $\Gamma_{ab}{}^c$ vanishes at that point.  This system is the local Minkowski space without gravity.  One way to see this is to look at the geodesic equation
\begin{equation}
\ddot{w}^a + \Gamma_{bc}^{a}w^bw^c = 0.
\end{equation}
If the connection $\Gamma_{bc}^{a}$ vanishes at a point then at least in a small neighborhood of that point a body behaves just like in Minkowski space free of gravity.

We want to remark that spacetime has become dynamical but it is still a stage for matter.  Matter is still moving \emph{on} spacetime.  The new feature of general relativity is that spacetime reacts to the presence of matter.  We will come back to this point later.  

\section{The equivalence principle,  again.}

In the previous section we have seen that the equivalence principle is fundamental to gravity.  Here we want to suggest that there is another way to implement the equivalence principle that does not rely on metrics.  To do so we have to make somewhat of a detour through the theory of many-body systems.  Let us begin with the simple example of an Ising model in one dimension.  Its Hamiltonian is given by
\begin{equation}
H = \sum_{i=1}^{N} \sigma_i\sigma_{i+1}.
\end{equation}
A possible ground state of this Hamiltonian is given by
\begin{equation}
\vert 0 \rangle = \vert 0 \ldots 0 \rangle.
\end{equation}
Excitations above the ground state have the form
\begin{equation}\label{eqn:spinwave}
\vert k \rangle = \sum_{n=1}^{N} \exp\left({2\pi i\; \frac{k n}{N}}\right)\ \vert n \rangle,
\end{equation}
with 
\begin{equation}
\vert n \rangle = \vert 0 \ldots 1 \ldots 0 \rangle,
\end{equation}
and the $1$ occurs at the $n$th position.  Note that the vacuum $\vert 0 \rangle$ is degenerate.  In this case the set of possible vacua $\mathcal{V}$ can be identified with the sphere $S^2$:
\begin{equation}
\mathcal{V} = S^2
\end{equation}
Each of the possible vacua has the same set of excitations.  In this case these are just the simple spin waves of equation (\ref{eqn:spinwave}).

Let is now imagine an internal observer in such a system.  By this we mean an observer who has only access to the emergent degrees of freedom,  i.e. the low-lying excitations and not the spins themselves.  These excitations are to the observer what elementary particles are to us.  The point that we want to stress now is that using only these excitations it is impossible for the observer to tell in what vacuum state she happens to be in.  All measurements are records of relations among excitations.  An excitation is distinguished by the way it interacts with other excitations.  Any point in $\mathcal{V}$ has the same set of excitations which in turn have the same interactions with each other.  Thus,  any point in $\mathcal{V}$ looks to an internal observer just like any other point in $\mathcal{V}$.  If, in particular, the excitations are such that the physics is as in Minkowski space then every vacuum in $\mathcal{V}$ gives rise to the same Minkowski space. 

It is here that we make the connection with the equivalence principle.  In gravity the equivalence principle states that locally we can not determine whether there is a gravitational field or not.  By putting ourselves in an Einstein elevator gravity disappears.  Locally there is no gravity.  Gravity is in the non-trivial effect that moving from one Minkowski space to another has.  We have seen that the vacua for many-body systems provide a similar structure.  If instead of having just one global vacuum we have a map
\begin{equation}
v :  M \longrightarrow \mathcal{V} 
\end{equation}
that assigns a (possibly) different vacuum to every point of spacetime $M$ then locally the physics would be indistinguishable from a constant vacuum but globally there are effects because the map $v$ is not constant.  We claim that these effects constitute gravity. 

Thus instead of saying that gravity arises because we are connecting different Minkowski spaces we are saying that gravity is the result of connecting different vacua. 

\section{The gravitational constant}

We want to explore this new view of gravity by showing how to calculate the gravitational constant $G$.  To do this let us distinguish between the gravitational mass $\mathsf{m}$ and the inertial mass $m$ of an object.  The gravitational mass $\mathsf{m}$ is defined by Newton's law of gravity.  Two objects with gravitational masses $\mathsf{m}_1$ and $\mathsf{m}_2$ attract each other with the force
\begin{equation}
F = \frac{\mathsf{m}_1\mathsf{m}_2}{r^2},
\end{equation}
where $r$ is the distance between the two masses.  This is not the way we usually write Newton's law.  Instead of using the gravitational masses $\mathsf{m_i}$,  $i=1,2$,  we use the inertial masses $m_i$, $i=1,2$.  Whereas the gravitational mass is about the gravitational force between two masses,  the inertial mass is about how hard it is to change the state of motion of one mass alone.  It is a remarkable experimental fact that there is a constant $C$ such that
\begin{equation}
m = C \mathsf{m}.
\end{equation}
We can thus write Newton's law of gravity in terms of the inertial masses:
\begin{equation}
F = G \frac{m_1 m_2}{r^2},
\end{equation}
where $G$ is Newton's gravitational constant.  Comparing the two expressions for Newton's law we see that 
\begin{equation}\label{eqn:gravconstant}
G = C^{-2}.
\end{equation}
To find Newton's constant we thus have to calculate the constant $C$ that relates the gravitational mass $\mathsf{m}$ to the inertial mass $m$ of an object.  

Let us begin by obtaining an expression for the inertial mass of an object (see \cite{dreyer} for more details).  For simplicity let us assume that $\mathcal{V}$ is just parametrized by one scalar parameter $\theta$.  Let us also assume that far away from any excitations the system is in the ground state $\theta_0$.  Excitations are then deviations $\delta\theta$ from the ground state $\theta_0$.  We can then go one step further and let these excitations form bound states.  It is the gravitational mass $\mathsf{m}$ of such a bound states that we want to determine.  

Away from the bound state the vacuum is approximately $\theta_0$.  The farther away we go from the bound state the better the approximation.  Next to the bound state though the situation has to be different.  Because the bound state is made up out of excitations and the excitations are themselves deviations from $\theta_0$ the vacuum next to the bound state can not be $\theta_0$.  As we will see it is this deviation that gives rise to gravity.  Let us assume that there is a term in the Hamiltonian of the form
\begin{equation}\label{eqn:hamiltonian}
\frac{1}{8\pi}\int d^3x\; (\nabla \theta)^2.
\end{equation}
(the factor of $1/8\pi$ is included here for future convenience).  A term like this tries to suppress deviations from $\theta_0$.  It also is responsible for the motion of the excitations.  Because of this term and because the presence of a bound state changes the vacuum there will be a force between any two bound states.  This is easily seen by analogy with electrostatics.  If we replace $\theta$ with $\phi$,  the electric potential,  equation (\ref{eqn:hamiltonian}) becomes the energy of the electric field.  From this energy we can obtain the force between two bound states by differentiating with respect to the position of one of the masses.  As in electrostatics we obtain the force
\begin{equation}
F = \frac{\mathsf{m}_1\mathsf{m}_2}{r^2},
\end{equation}
if we define the masses (i.e. the charges) by
\begin{equation}\label{eqn:gravitational}
\mathsf{m}_i = \frac{1}{4\pi}\int_{\partial \mathsf{m}_i} d\sigma \cdot \nabla\theta,\ \ i=1,2.
\end{equation}
Here $\partial \mathsf{m}_i$ is the boundary of the bound state $i$.  

To find the inertial mass we now make an assumption about the momentum of a moving mass.  We do this by taking the analogy with electromagnetism one step further.  The momentum of a moving spherical shell with radius $a$ and charge $q$ is given by \cite{singal}
\begin{equation}
p = \frac{2}{3}\frac{q^2}{a}\gamma v,
\end{equation}
where $\gamma= \sqrt{1 + v^2}$.  For small velocities $v$ the factor $\gamma$ is roughly unity and we obtain
\begin{equation}
p = \frac{2}{3}\frac{q^2}{a} v.
\end{equation}
We now assume that our system behaves in the same way as electrodynamics.   Thus we assume that the momentum of the moving mass is given by
\begin{equation}
p = \frac{2}{3} \frac{\mathsf{m}^2}{a}v
\end{equation}
Here $a$ is now the linear extension of the bound state that we are looking at.  To read off the expression for the inertial mass we rewrite this equation as follows:
\begin{equation}
p = \frac{2}{3} \frac{\mathsf{m}}{a}\; \mathsf{m} v
\end{equation}
The whole expression multiplying the velocity $v$ is equal to the inertial mass $m$.  Thus 
\begin{equation}\label{eqn:inertial}
m = \frac{\mathsf{2 m}}{3 a}\; \mathsf{m},
\end{equation}
and the constant $C$ that relates the inertial and the gravitational mass is given by 
\begin{equation}
C = \frac{\mathsf{2 m}}{3 a}.
\end{equation}
From equation (\ref{eqn:gravconstant}) we then find the gravitational constant $G$:
\begin{equation}
G = \left(\frac{3 a}{\mathsf{2 m}}\right)^2
\end{equation}
The strange factors of two and three appear here because we used the expression for the momentum of a charged shell.  The details differ if another charge distribution is used instead but are always of order unity. 

\section{Discussion and outlook}

What is quantum gravity?  This is the question that started this paper.  The usual answer is that quantum gravity it the quantization of general relativity,  the classical theory of gravity.  Quantum gravity is thus a quantum theory of spacetimes.  The technical question that one has to tackle is how to construct a theory of superpositions of spacetimes.  This has proven difficult for conceptual as well as mathematical reasons.  We have argued here that quantum gravity is very different from a theory of quantized metrics.  First we have argued that gravity is not part of the fundamental theory but instead is an emergent feature of the low-energy theory.  Then we have argued that gravity is not an excitation of the theory but a non-perturbative feature.  Gravity is due to the spatially changing vacuum of the theory.

People have looked at emergent theories of gravity for a while (see \cite{lorenzo} for a recent review).  Our approach is new in that we do not aim at deriving Einstein's equation from our setup.  Instead of arguing that gravity is about metrics and that the aim of an emergent theory of gravity should be to derive Einstein's theory as the low-energy emergent theory we say that gravity is not about metrics at all.  Instead it is about how the vacuum reacts to the presence of excitations of this vacuum.  This is the really new feature of our theory.  No more spacetimes.  

One appealing aspect of our setup that is,  that there is no need for a quantization of the theory.  Since we start from a quantum theory,  no quantization required.  This is much like theories in solid state physics.  Instead of starting with classical theory that needs to be quantized,  one usually starts from a theory expressed in terms of simple quantum systems instead.  Together with a Hamiltonian expressing the interaction of these quantum systems,  this provides the complete description of the system.  The hard part is then is then to find the low-energy behavior of this theory.  Quantization never enters the picture. 

We think that this relation between the classical limit and the underlying quantum theory is as it should be.  The idea that the more fundamental quantum theory is the quantization of the emergent classical theory always struck us as an unlikely scenario.

We have shown how this new view of gravity can be used to calculate the gravitational constant $G$ from the underlying theory.  We did this by deriving expressions for the gravitational and the inertial mass of an object.  It is here that most of the questions about the theory arise.  How much of an assumption about the underlying theory is made in equations (\ref{eqn:gravitational}) and (\ref{eqn:inertial})?  Also more work is required in analyzing what the exact status of the equivalence principle is in our theory.  We have motivated our approach by appealing to the equivalence principle but it seems that it is conceivable that the expression that we have found for the gravitational constant is not universal and might vary from one particle species to another.  It also might be dependent on the energy of the particles.  All these questions will be analyzed in future work.

\section*{Acknowledgements}
I would like to thank Seth Lloyd for all the help he has provided both scientifically and  privately.  Without it this paper would not exist.  Similarly I would like to thank Achim Kempf for his support at the University of Waterloo,  where important parts of this paper were created.  Special thanks go to the Foundational Questions Institute,  FQXi,  for financial support and for creating an environment where it is alright to play with unorthodox ideas.  I would also like to thank Fotini Markopoulou for continuing discussions during the creation of this paper.


\begin{thebibliography}{WW}
\bibitem{pullin} J. Pullin, \emph{A First Course in Loop Quantum Gravity},  Cambridge University Press,  2011.
\bibitem{thieman}T. Thiemann,  \emph{Modern Canonical Quantum General Relativity},  Cambridge University Press,  2008.
\bibitem{rovelli}C. Rovelli,  \emph{Quantum Gravity},  Cambridge University Press,  2004.
\bibitem{loll} J. Ambjorn, J. Jurkiewicz,  and  R.  Loll,  \emph{Quantum gravity as sum over spacetimes},  arXiv:0906.3947.
\bibitem{dreyer} O. Dreyer, \emph{Why things fall},  in proceedings of "From Quantum to Emergent Gravity: Theory and Phenomenology",  PoS(QG-Ph)016,  2007.
\bibitem{singal} A. K. Singal,  \emph{Energy-momentum of the self-fields of a moving charge in classical electromagnetism}, Journal of Physics A (25)1992, p. 1605
\bibitem{lorenzo} L. Sindoni,  \emph{Emergent models for gravity: an overview
},  arXiv:1110.0686.
\end{thebibliography}
\end{document}